\documentstyle[aps,prl,preprint]{revtex}
\tightenlines
\begin{document}
\draft
\preprint{}
\title{The Structure of the Vortex Liquid at the Surface of a Layered
Superconductor}
\author{Andreas Kr\"amer}
\address{Dept.~of Applied Physics, Stanford University, Stanford, CA 94305,
USA}
\author{J.~Enrique D\'\i az-Herrera}
\address{Dept.~de Fisica, Univ.~Autonoma Metropolitana, Mexico City,
Mexico}
\date{\today}
\maketitle
\begin{abstract} 
A density-functional approach is used to calculate 
the inhomogeneous vortex density distribution in the flux liquid phase at the 
planar surface of a layered 
superconductor, where the external
magnetic field is perpendicular to the superconducting layers and parallel
to the surface. The interactions with image vortices are treated within 
a mean field approximation as a {\em functional} of the vortex density.
Near the freezing transition strong vortex density fluctuations are found to 
persist far into the bulk liquid. 
We also calculate the height of the Bean-Livingston surface barrier.
\end{abstract}
Fluctuations of vortices in high-$T_c$ superconductors give rise to 
a number of 
novel effects \cite{Blatter,Brandt} the most prominent of which is probably the
melting of the vortex lattice into a vortex liquid which was first 
proposed by Nelson \cite{Nelson} in 1988. While much work has been done
to study the bulk 
properties of the vortex system, 
the influence of fluctuations on the behavior of flux vortices at the 
surface of a superconductor only rarely has been taken into account.
Recently Brandt, Mints, and Snapiro \cite{Brandt1} calculated the 
interaction of a single stack of thermally fluctuating pan-cake vortices 
with the planar surface of a layered superconductor.  They found  that 
fluctuations shift the short-range 
attraction between the vortices and their mirror-images to a long-range
dipole-dipole interaction. In this Letter we consider a related 
problem, namely the interaction of a vortex-{\em liquid} with a planar surface.

In order to describe the vortex system in the liquid regime we will apply the 
density-functional theory of classical liquids \cite{Lehrbuch}. The theory 
for the
bulk vortex liquid has been worked out by Sengupta {\em et al.} \cite{Sengupta}
who have calculated bulk correlation functions and treated the freezing of 
the vortex
liquid into the solid phase. In the present work we will compute the 
inhomogeneous density distribution
of the vortices in front of a planar surface using the 
Wertheim-Lovett-Mou-Buff (WLMB) equation \cite{Wertheim}
together with a mean-field treatment of the image-forces.\cite{Patey} 
The same method has recently successfully been applied in a different context, 
namely the structure of liquid water at a metallic interface \cite{water}.
From the equilibrium vortex-density profiles at the surface 
we calculate the local magnetic field which together with the image 
interactions determines the potential energy barrier \cite{Bean}
for pan-cake vortices entering into the sample. From this the location of the 
irreversibility line
(IL) in the $B$-$T$-diagram is estimated for the case of zero bulk pinning
and vanishing {\em geometrical} surface barrier \cite{Mayer}. 
 
Before discussing  the surface problem we will shortly reconsider the density 
functional approach for the bulk vortex liquid following Sengupta {\em et al.}
\cite{Sengupta}.
This enables us to compute the bulk direct correlation functions we need
as input into the surface calculations described below.  In the limit of zero 
interlayer Josephson
coupling and for ${\bf H}\parallel {\bf c}$ the vortices in an $N$-layer 
superconductor can be viewed as a 
{\em two dimensional} classical system of $N$ species of point particles 
interacting via the (asymptotically) logarithmic pair potential $V_n(r)$ of
two vortices separated by $n$ layers. 
In the limit $N\rightarrow\infty$ the 
Fourier transform of this  pair potential is given by \cite{Feigelman}
\begin{equation}
\label{Fourier}
V({\bf k})=\frac{\epsilon_0\lambda^2\left[k_\perp^2+(4/d^2)\sin^2(k_zd/2)
\right]}
{k_\perp^2\left[1+\lambda^2k_\perp^2+(4\lambda^2/d^2)\sin^2(k_zd/2)\right]},
\end{equation}
where $k_z$($k_\perp$) is the component of the wave-vector ${\bf k}$ 
perpendicular
(parallel) to the layers.
The parameter  $\epsilon_0=d\Phi_0^2/4\pi\lambda^2$ is the
characteristic energy scale, $\beta$ is the inverse temperature, 
$d\ll\lambda$ is the distance of the layers, and 
$\Phi_0$ the flux quantum. The (2D-) pair potentials are repulsive for pancakes
residing in the same layer and attractive but by the factor $(d/2\lambda)
\exp(-nd/\lambda)$ weaker for pancakes in two separate layers of distance 
$nd$. 
However, the attractive and repulsive logarithmic potentials compensate
each other on distances larger than $\lambda$ leading to the usual short-range
repulsion ($\sim K_0(r/\lambda)$) of Abrikosov vortices between
two straight stacks of pancakes.

From the pair potentials $V_n(r)$ one can derive the liquid-state pair 
correlation functions
$g_n(r)$ within standard liquid theory. As in Ref.~\cite{Sengupta} we use 
the hypernetted chain (HNC) approximation. The basic equations are then the
Ornstein-Zernicke equation 
\begin{equation}
\label{OZ}
h_n(r)=c_n(r)+\varrho_B\sum_m\int d^2{\bf r'} h_m(r')c_{n-m}
(|{\bf r}-{\bf r'}|)
\end{equation}
and the HNC-closure relation
\begin{equation}
\label{closure}
g_n(r)=\exp\left[-\beta V_n(r)+h_n(r)-c_n(r)\right]
\end{equation}
for the total correlation function $h_n(r)=g_n(r)-1$ and the direct
correlation function $c_n(r)$, where $\varrho_B$ is the bulk 
density of the
vortex liquid. For the numerical (iterative) solution of these equations it
is appropriate to reformulate Eq.~(\ref{OZ}) in ${\bf k}$-space and to
separate the direct correlation function $c_n(r)$ into a long-range and a 
short-range part. (Note that $c_n(r)\rightarrow -\beta V_n(r)$ as $r\rightarrow
\infty$.) We were able to obtain
self consistent solutions of the complete set of Eqs.~(\ref{OZ}) and 
(\ref{closure})
for a periodic system with 512 layers using a Fast Hankel transform 
method\cite{Siegman}. 

The phase boundary between the vortex liquid and the
vortex solid was obtained in a different way than in Ref.~\cite{Sengupta} 
namely 
by examining the thermodynamical stability of the system, i.e.~we looked
at the spinodal curve rather than the coexistence line at the first order 
phase transition. The system becomes unstable if the second functional
derivative of the grand canonical potential w.r.t.~the density, 
${\cal M}({\bf r},z,{\bf r'},z')=\delta^2\Omega/
\delta\varrho({\bf r},z)\delta\varrho({\bf r'},z')$, is no longer positiv 
definite.
\cite{Kasch} In the case considered here ${\cal M}({\bf r},z,{\bf r'},z')$ is 
diagonal in 
${\bf k}$-space because of the translational invariance of the vortex-system 
in all three space directions, ${\cal M}({\bf k},{\bf k'})=(1-c(k_z,k_\perp))
\delta({\bf k}-{\bf k'})$. As expected, near the spinodal curve  the system 
becomes 
unstable w.r.t.~to fluctuations of the density $\sim e^{i{\bf k}_\perp{\bf r}}$
with $|{\bf k}_\perp|\approx 2\pi\varrho_B^{1/2}$ ($k_z=0$), which indicates
that the formation of a vortex lattice becomes favorable. The freezing line 
(spinodal curve) in the ($B$,$T$)-diagram is then obtained by 
extrapolating $1-c(k_z,k_\perp)$ to zero. For high magnetic fields $B$
the transition temperature approaches the 2D-melting temperature $T_{2D}$
which is independent of the vortex density \cite{Sengupta}. 

We now turn to the calculation of the
local vortex density near the surface of the superconductor.
We start with the
Wertheim-Lovett-Mou-Buff (WLMB) equation \cite{Wertheim}  which is an exact 
relation 
following from classical density-functional theory. 
In a more general context this system of integro-differential equations
relates 
the particle densities $\varrho_\alpha$ of species $\alpha$ to an external 
potential
$V_\alpha$: 
\begin{equation}
\label{WLMB0}
\nabla_1\log\varrho_\alpha(1)=-\beta\nabla_1 V_\alpha(1)
+\sum_{\beta}\int d2\nabla_2\varrho_\beta(2)
c_{\alpha\beta}(1,2,[\varrho])
\end{equation}
 Here, $c_{\alpha\beta}(1,2,[\varrho])$ is the 
{\em inhomogeneous} direct correlation function which is a functional of
the density itself. 
In the following we will again work in the HNC
approximation and replace $c_{\alpha\beta}(1,2,[\varrho])$ by its bulk version.
We introduce a cartesean coordinate system where the ($y$,$z$)-plane is the 
surface of the superconductor which fills the half-space at $x>0$. Because of 
the translational invariance
in $z$-direction we are left with
one equation for the vortex density $\varrho(x)$ in each layer which is a 
function of $x$ only.
Now, the bulk direct correlation function $c_n(r)$ is written as a sum of a 
long-range and a short-range part, 
$c_n(r)=c_n^{SR}(r)-\beta V_n(r)$, we carry out
the integration parallel to the surface, and define
\[
c^{SR,\parallel}(x)=\sum\limits_n\int\limits_x^\infty
\frac{2rdr}{\sqrt{r^2-x^2}}
c_n^{SR}(r).
\]
With $\sum_n V_n(r)=(\epsilon_0/2\pi)K_0(r/\lambda)$
the WLMB-equation (\ref{WLMB0}) then becomes 
\begin{equation} 
\label{WLMB2}
\frac{d}{dx}\log\varrho(x)=
\beta\Delta\varrho\lambda\epsilon_0 e^{-x/\lambda}
+ \beta F^I(x,[\varrho])
+\int\limits_0^\infty dx'\frac{d\varrho}{dx'}c^{SR,\parallel}(|x-x'|)
-\frac{\beta\epsilon_0\lambda}{2}\int\limits_0^\infty dx'\frac{d\varrho}{dx'}
e^{-|x-x'|/\lambda}.
\end{equation}
Eq.~(\ref{WLMB2}) is the basis for our numerical calculation.
The first term on the r.h.s. of Eq.~(\ref{WLMB2}) gives the force produced
by an exponentially decaying shielding current at the surface of the 
superconductor which arises from the homogeneous solution of the London 
equation.
The second term is the image force 
which is the sum of the
self-image force $F^{SI}$  and the force $F^{OI}$ produced by the images of all
other pancakes, $F^{I}(x,[\varrho])=F^{SI}(x)+F^{OI}(x,[\varrho])$. 
The third term in Eq.~(\ref{WLMB2}) gives the mean force caused by the 
liquid correlations and the last term represents the electromagnetic 
vortex-vortex interaction.

Eq.~(\ref{WLMB2}) has to be solved for the density $\varrho(x)$ at $x>0$ 
with $\varrho(x)\stackrel{x\rightarrow\infty}{\longrightarrow}\varrho_B$
 while the parameter $\Delta\varrho$  is 
determined by the condition that the 
{\em potential of mean force}, $V_{\mbox{\footnotesize{PMF}}}(x)=
-k_BT\ln(\varrho(x)/\varrho_B)$ is zero
at the surface ($\varrho(0)=\varrho_B$). 
The local magnetic field
 $B(x)$ then reads
\begin{equation}
\label{field}
B(x)=\Phi_0(\Delta\varrho-\varrho_B) e^{-x/\lambda}
+\frac{\Phi_0}{2\lambda}\int\limits_0^\infty\varrho(x')
e^{-|x'-x|/\lambda}dx',
\end{equation}
the applied field is $B_a=B(0)$,
and the bulk magnetic field is given by 
$B_0=B(\infty)=\varrho_B\Phi_0$.

In the following we will derive an expression for the image forces $F^I$ 
within a
mean field approximation.
The self image force acting on a pancake vortex at position ${\bf r}=(x,0)$ 
is given by
\begin{equation}
\label{self}
F^{SI}(x)=-\frac{\epsilon_0}{4\pi x}\;\;\;(x>\xi),
\end{equation}
where the cut-off $\xi$ is related to the coherence length.
The $x$-component of the force exerted on the same vortex by the image of a 
second  vortex located
at ${\bf r'}$ in layer $n$ is  
\[
f_n({\bf r},{\bf r'})=\frac{V'_n\left(\sqrt{r_{12}^2+4xx'}\right)(x+x')}
{\sqrt{r_{12}^2+4xx'}}
\]
with $r_{12}=|{\bf r}-{\bf r'}|$.
Thus, the average total force produced by the non-self images can be written
\begin{equation}
\label{other}
F^{OI}(x,[\varrho])=\sum\limits_n\int f_n({\bf r},{\bf r'})
\varrho({\bf r'})h_n({\bf r'}|{\bf r}) d{\bf r'}
\end{equation}
where $h_n({\bf r'}|{\bf r})$ is the total correlation function for the 
inhomogeneous liquid distribution at the 
surface.
Here we assume that forces {\em parallel} to the surface cancel each other on 
the average.  
If the total correlation function $h_n({\bf r'}|{\bf r})$ is again 
approximated 
by its bulk version $h_n(r_{12})$ the non-self image force $F^{OI}$ can 
finally be expressed as a {\em linear functional} of the vortex density 
$\varrho(x)$,
\begin{equation}
\label{OI}
F^{OI}(x,[\varrho])=\int\limits_0^\infty\varrho(x'){\cal J}(x,x')dx'.
\end{equation}
The kernel ${\cal J}(x,x')$ is given by
\begin{equation}
\label{kernel}
{\cal J}(x,x')=(x+x')\sum\limits_n
\int\limits_{|x-x'|}^\infty\frac{2r_{12}h_n(r_{12})V'_n\left(
\sqrt{r_{12}^2+4xx'}\right)}{\sqrt{r_{12}^2-(x-x')^2}\sqrt{r_{12}^2+4xx'}}
dr_{12}.
\end{equation}
It is important to notice that the force of the non-self images cancels the 
long-range self-image force for sufficient large $x$. Indeed, writing 
$V_n(x)\sim q_n \ln (x/\xi)$ for $x>\lambda$ with `charges' $q_0=-1$, 
$q_{n\neq 0}\approx
(d/2\lambda) e^{-nd/\lambda}$ ($\sum_nq_n=0$),
setting $\varrho(x)=\varrho_B$, and
using the `screening condition' for the total correlation function $h_n(r)$,
$q_0+\sum_n\int d^2{\bf r}\varrho_Bq_nh_n(r)=0$, one can perform the integrals
(\ref{OI}) and (\ref{kernel}) analytically and gets 
$F^{OI}\stackrel{x\rightarrow\infty}
{\longrightarrow} \epsilon_0/4\pi x=-F^{SI}$.

In what follows we present results of the numerical solution of
Eq.~(\ref{WLMB2}). We introduce the dimensionless parameters 
$f=\lambda^2\varrho_B$,
$\Gamma=\beta\epsilon_0$, $\delta=d/\lambda$, and $\kappa=\lambda/\xi$. All 
calculations have been performed for
$\kappa=100$ and $\delta=0.01$ while $f$ and $\Gamma$ were varied.
About $10^4$ iterations of Eq.~(\ref{WLMB2}) on a grid of 2048 sites were 
needed  
to achieve convergence. However, when approaching the freezing transition
of the vortex liquid the rate of convergence becomes slower and the numerical
procedure finally diverges. The kernel ${\cal J}(x,x')$ for the image-forces
(Eq.~(\ref{kernel})) can be calculated once and then stored in a lookup-table.

Fig.~1 shows the so-obtained vortex density profiles for $f=0.75$ and
parameters $\Gamma=100$, 250, and 360. Freezing occurs at approximately
$\Gamma=380$. It is seen that near the freezing transition 
strong density fluctuations persist far into the bulk liquid, so that
the vortex-liquid is still solid-like at the surface. These
fluctuations are washed out at higher temperatures (lower values of $\Gamma$).
 The corresponding
magnetic field near the surface is shown in the left-hand inset  of Fig.~1.
Directly at the surface (invisible in Fig.~1) the density rapidly drops from
$\varrho_B$
to almost zero, which reflects the existence of a Bean-Livingston surface 
barrier \cite{Bean}.
This surface barrier is shown in the right-hand inset of Fig.~1 where the
potential of mean force $V_{\mbox{\footnotesize{PMF}}}$ is plotted for several
values of the parameter $f=0.6,0.75,1.0,2.0$ and $\Gamma=250$. The barrier 
height $U$ turns out
to be nearly independent of the temperature but decreases  with increasing 
magnetic field (parameter $f$). This is shown in the inset of Fig.~2. 

When the applied field is increased above its equilibrium value vortices will
start to enter into the sample, where the flux-entry rate ${\cal R}$ is 
determined by the Boltzmann factor $e^{-\beta U}$.
The irreversibility line (IL) separates the irreversible and the reversible 
part of the phase diagram on experimental time scales and is thus given by
${\cal R}=const.$ We can therefore estimate 
(up to a constant factor) the location of the surface-barrier induced IL in
the $B$-$T$-diagram by the condition $U(B)=ak_BT$, where $a$ is a constant.

In order to compare with experimental results
we will in 
the following use typical parameters for Bi$_2$Sr$_2$CaCu$_2$O$_{8}$ (BSCCO).
The London penetration depth is taken to have a temperature dependence 
 $\lambda(T)=\lambda(0)(1-(T/T_C)^4)^{1/2}$ with $\lambda(0)=1400$ $\AA$ and
$T_C=93$ K, $\kappa=100$, and
$d=15$ $\AA$. Fig.~2 (broken line) shows the melting line (spinodal curve) of 
the flux-line lattice in
BSCCO as obtained by the method described above. For $T\gtrsim 50$ K it 
agrees well with the experimental data of Majer {\em et al.} \cite{Mayer} 
($\diamond$) . 
Deviations
at low temperatures are possibly an effect of the Josephson interaction 
which has not been taken into account here.
Fig.~2 (solid line) also shows the estimated location of the IL. A
constant factor $a=42$ was chosen such that the curve lies near the 
experimentally obtained IL ($+$) from Ref.~\cite{Mayer}
for a sample of triangular cross-section, where the geometrical surface 
barrier is 
absent. Since bulk pinning plays no role in these experiments 
either, as argued in
Ref.~\cite{Mayer}, the IL shown in Fig.~2 originates from a 
{\em Bean-Livingston}
surface barrier alone. Though our simple calculation may not be
applicable to this experiment in detail because of the more complicated
sample geometry and possible effects of the surface roughness, the
qualitative features
of the IL found here are the same as those found in the experiment.
In both cases the IL 
has  a negative slope $dB/dT$ and is  
much steeper than the melting line crossing the latter one at some
point.

In conclusion we have calculated the density distribution of pan-cake vortices
in the flux liquid phase at the surface of a layered superconductor using 
density functional theory and a mean field treatment of the image-interactions.
Near the melting line strong density fluctuations are found to reach far
into the bulk liquid.
This could possibly enhance bulk-pinning near the surface. It is shown that
a Bean-Livingston barrier exists from which we estimate the location of the 
irreversibility line (IL) in the $B$-$T$-plane which is qualitatively 
consistent
with experimental findings for BSCCO crystals.

One of us (A.K.) thanks the Deutsche Forschungsgemeinschaft for financial 
support.

\begin{figure}
\caption{Density profile $\varrho(x)$ for constant bulk density ($f=0.75$) and 
different temperatures ($\Gamma=100$, 250, and 360).
 The right-hand inset shows the potential of
mean force (PMF) at constant temperature ($\Gamma=250$) for different bulk 
densities
$(f=0.6, 0.75, 1.0, 2.0$, from the upper to the lower curve). The left-hand 
inset gives the internal magnetic field $B(x)$ at the surface for the same 
set of parameters.}
\end{figure}

\begin{figure}
\caption{The melting line of the vortex lattice in Bi$_2$Sr$_2$CaCu$_2$O$_{8}$:
Experimental data from Ref.~[11]
($\diamond$) and this work (dashed line).
Irreversibility line (IL): Experimental data for a sample with triangular 
cross-section from Ref.~[11] ($+$) and
as estimated in this work (solid line, see text). The inset shows the
barrier height $U/\epsilon_0$ as function of $f$.}
\end{figure}

\end{document}